\documentclass[aps,cpb,reprint,twocolumn,superscriptaddress,showpacs]{revtex4-1}
\usepackage{graphicx}
\usepackage{mathrsfs}
\usepackage{bm}
\usepackage{amsmath}
\usepackage{dcolumn}
\usepackage{epstopdf}
\usepackage{dsfont}
\usepackage{amssymb}
\usepackage{tabularx}
\usepackage{array}
\usepackage{float}
\usepackage{color}
\usepackage{epstopdf}
\usepackage[colorlinks, linkcolor=blue,anchorcolor=blue,citecolor=blue,urlcolor=blue]{hyperref}
\usepackage{xcolor}
\usepackage{ulem}
\newcommand{\tr}{\mathrm{Tr}}

\usepackage{soul}
\setstcolor{red} 
\begin{document}

\title{Entanglement and energy transportation in the central-spin quantum battery}

\author{Fan Liu}
\affiliation{School of Physics, Northwest University, Xi'an 710127, China}

    \author{Hui-Yu Yang}
    \affiliation{School of Physics, Northwest University, Xi'an 710127, China}

    \author{Shuai-Li Wang}
    \affiliation{School of Physics, Northwest University, Xi'an 710127, China}

    \author{Jun-Zhong Wang}
    \affiliation{School of Physics, Northwest University, Xi'an 710127, China}

    \author{Kun Zhang}
    \email{kunzhang@nwu.edu.cn}
    \affiliation{School of Physics, Northwest University, Xi'an 710127, China}
    \affiliation{Shaanxi Key Laboratory for Theoretical Physics Frontiers, Xi'an 710127, China}
    \affiliation{Peng Huanwu Center for Fundamental Theory, Xi'an 710127, China}

    \author{Xiao-Hui Wang}
    \email{xhwang@nwu.edu.cn}
    \affiliation{School of Physics, Northwest University, Xi'an 710127, China}
    \affiliation{Shaanxi Key Laboratory for Theoretical Physics Frontiers, Xi'an 710127, China}
    \affiliation{Peng Huanwu Center for Fundamental Theory, Xi'an 710127, China}

  \begin{abstract}

     Quantum battery exploits the principle of quantum mechanics to transport and store energy. We study the energy transportation of the central-spin quantum battery, which is composed of $N_b$ spins serving as the battery cells, and surrounded by $N_c$ spins serving as the charger cells. We apply the invariant subspace method to solve the dynamics of the central-spin battery with a large number of spins. We establish a universal inverse relationship between the battery capacity and the battery-charger entanglement, which persists in any size of the battery and charger cells. Moreover, we find that when $N_b=N_c$, the central-spin battery has the optimal energy transportation, corresponding to the minimal battery-charger entanglement. Surprisingly, the central-spin battery has a uniform energy transportation behaviors in certain battery-charger scales. Our results reveal a nonmonotonic relationship between the battery-charger size and the energy transportation efficiency, which may provide more insights on designing other types of quantum batteries. \\

        \vspace{0mm}


        \vspace{0mm}

  \end{abstract}
	
\maketitle

\section{Introduction}

    Quantum technology has demonstrated promising advantages in various of fields, including computing, communication, and simulation.\textsuperscript{\cite{Nielsen,Bennett93,Gisin02,Ladd10, Reiher17, Rudinger22, Duan01}}
    Quantum resources, such as coherence and entanglement, are essential in many quantum protocols as well as simulating the physical models.\textsuperscript{\cite{Koepsell21, Muniz20, Blatt12, Tamura20, Niu21, Guo21}}
    For example, quantum resources provide substantial benefits in energy manipulation, namely the quantum battery.\textsuperscript{\cite{Campisi11, Yang23, shi20, Ji22, Dou22, Dou22A, Wang20, Lu21, Uzdin15, F13, Alicki13,Barrios17, Altintas14, Park13, Seah21, Manzano18, Goold16, Strasberg17, Watanabe17, Zhang19}}
    Quantum battery based on the organic microcavity has been realized in experiments, showing great advantages on energy transportation.\textsuperscript{\cite{Quach22}}

    The primary goal of quantum battery research is to explore how to increase the energy storage capacity and/or maximize the speed of charging.\textsuperscript{\cite{Ferrar18, Fusco16,Binder15, Andolina18, Santos21, Rossini20,Yu23,Yang231}} For example, the quantum battery has the maximal stored energy in the steady state if the battery and the charger have the same coupling strength with a shared common bath.\textsuperscript{\cite{Yao21}} The Dicke quantum battery (based on the Dicke model) with $N_b$ battery cells can achieve a superextensive charging rates, namely the charging power $P\propto N_{b}^{3/2}$, for different initial states.\textsuperscript{\cite{Andolina19}} The fast-charging advantage for the quantum battery comes from the coherent cooperative interaction between the charger and the battery.\textsuperscript{\cite{Xiang23}}

    Traditional batteries use an electric field to store energy, which is then transformed into electricity by a redox process. While quantum batteries can take advantage of quantum degrees of freedom, such as spin, to store and transport energy.\textsuperscript{\cite{Jian07, Bortz20}} Consider $N_{b}$ noninteracting spins as the battery cells and $N_{c}$ noninteracting spins as the charger cells. The battery and charger spins allow interactions, which is required for charging. See Fig. \ref{fig1}. The above construction is called the central-spin model, or central-spin battery in our study.\textsuperscript{\cite{Gaudin76,Gaudin22}} The central-spin model appears in various nanostructures, such as semiconductors, quantum dots, carbon nanotubes, and nitrogen-vacancy centers in diamond.\textsuperscript{\cite{Faribault19, Doherty13, JY23, Li20}} Specifically, the hyperfine interaction between the spin of electrons in quantum dots and the spin of surrounding nuclei can be well described by the central-spin model.\textsuperscript{\cite{Schliemann02, Khaetskii02, Deng06}}

    Previous study has clarified the charging power of the central-spin battery. When the number of chargers is less than the number of batteries, we have the charging power $P\propto {N_{b}}^{1/2}$; when the number of chargers is much larger than the number of batteries, we have the charging speedup, namely $P\propto {N_{b}}^{3/2}$.\textsuperscript{\cite{Peng21}} Although the central-spin battery has a charging advantage due the coherent collective interaction, the entanglement between the charger and the battery or inside the battery spins may prohibit the battery achieving the maximal energy storage.\textsuperscript{\cite{Kamin21, Hovhannisyan13, Liu21, Shi22}} The question on how to realize the optimal energy storage in central-spin battery has never been addressed.

    In this work, we aim to explore when the central-spin battery can achieve the maximal energy storage. We focus on how the size of battery and charger, namely $N_b$ and $N_c$, influence the maximal energy storage of the central-spin battery. We also calculate the entanglement between the charger and the battery, which clarifies the role of quantum resource in quantum batteries. Although central-spin model is integrable,\textsuperscript{\cite{Gaudin76,Gaudin22}} traditional method such as Bethe ansatz can not capture the nonequilibrium dynamics of the integrable systems. Instead, we apply a invariant subspace method, which allows us to numerically characterize the dynamics of the system with large number of spins. Moreover, we establish the analytical results when $N_b = 2$ or $N_c = 2$.

    The paper is organized as follows. In Sec. \ref{S-CB}, we introduce the central-spin model as well as the the invariant subspace method. In Sec. \ref{S-EAET}, we analytically study the battery capacity and the battery-charger entanglement with $N_b =2$ or $N_c=2$. In Sec. \ref{S-COET}, we focus on the conjectured resonant condition, namely $N_b = N_c$, and numerically study the battery capacity and the battery-charger entanglement. Conclusion and outlook are presented in the final section.

\section{Central-spin battery}\label{S-CB}

    \begin{figure}[t]
        \includegraphics[width=\columnwidth]{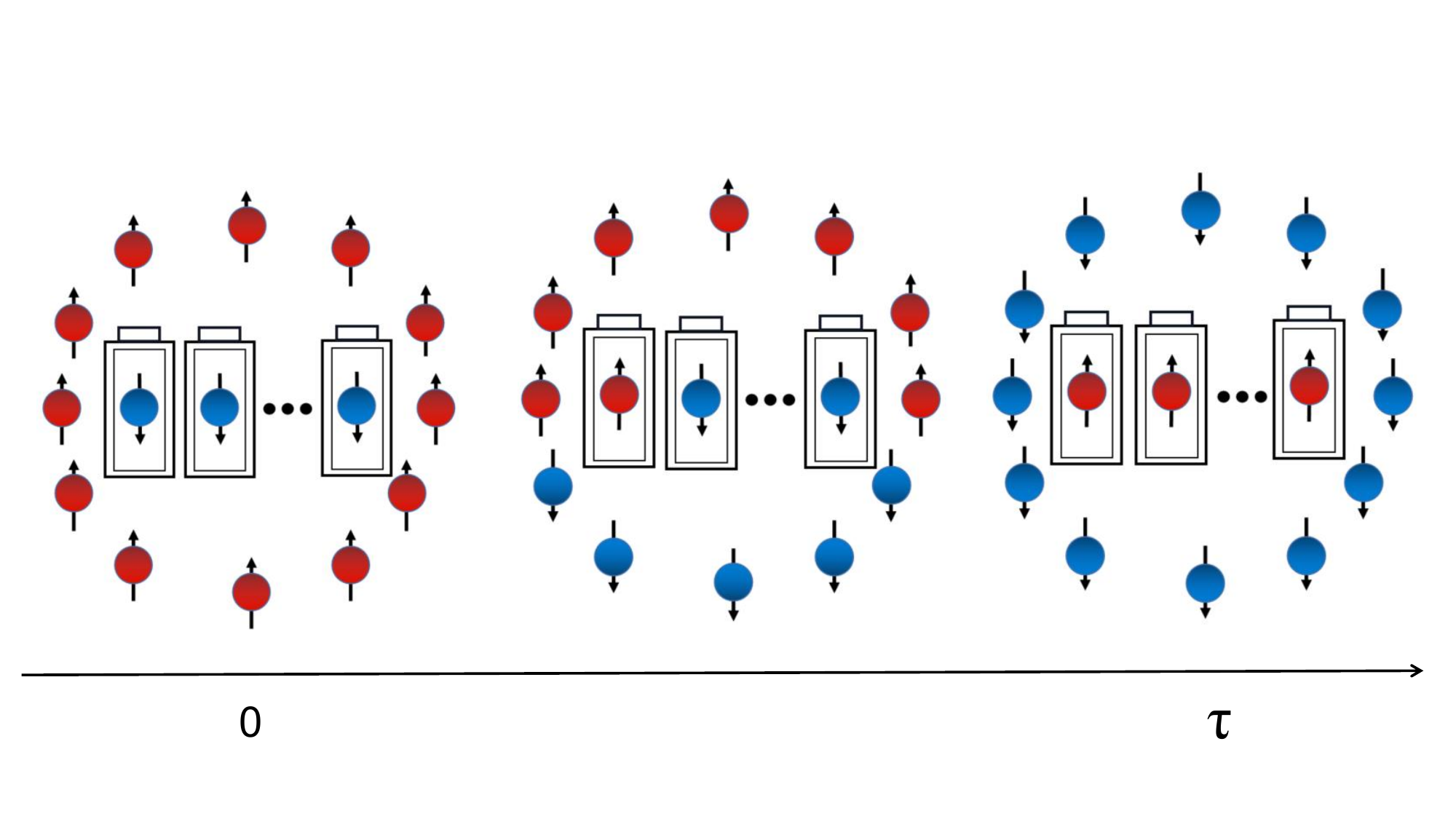}
        \caption{Schematic diagram of the central-spin battery. The spins in the center serve as the battery. The surrounding spins work as the charger. Initially, the battery spins are in the ground states (spin down with the blue color) and the charger spins are in the excited states (spin up in the red color). In the time interval $0<t<\tau$, the interaction Hamiltonian $H_{I}$ in Eq. (\ref{eq:H_b}) is turned on and the battery is charging. (That is, the flip-flop interaction and the Ising-type interaction between the battery and charger spins are turned on). Finally, when $t>\tau$, the interaction is turned off, and the energy is stored in battery.
        }\label{fig1}
    \end{figure}

    Consider the central-spin battery, which has the Hamiltonian
        \begin{equation}
        \label{CSB}
            H= H_{b}+H_{c}+ g(t) H_{I},
        \end{equation}
    composed with the battery Hamiltonian $H_{b}$, charger Hamiltonian $H_{c}$, and their interaction $H_{I}$. Specifically, we have
        \begin{subequations}
        \begin{align}
            \label{eq:H_b}H_{b}=&\omega_b S^z, \\
            \label{eq:H_c}H_{c}=&\omega_c J^z, \\
            \label{eq:H_I}H_{I}=&\lambda(S^+J^-+S^-J^+)+2\Delta S^z J^z.
        \end{align}
        \end{subequations}
    Here $S^{\alpha}=\sum_{j=1}^{N_b}\sigma_j^\alpha/2$ and $J^{\alpha}=\sum_{k=1}^{N_c}\sigma_k^\alpha/2$ with $\alpha=x,y,z$ are the total spin operators for the battery and charger spins respectively. Correspondingly, we have the spin ladder operators $J^\pm=J^x\pm iJ^y$ and the $S^\pm=S^x\pm iS^y$. The parameters $\omega_b$ and $\omega_c$ characterize the onsite energy of the battery and charger spins, corresponding to the strengths of the external magnetic field. The parameter $\lambda$ and $\Delta$ characterize the flip-flop interaction (spin XX-YY interaction) and the Ising-type interaction (spin ZZ interaction), respectively. Due to the flip-flop interaction, the battery spins will be excited to high-energy states at the cost of decreasing the number of spin-up charging units. The charging process is accomplished by turning on the $H_I$ interaction between the battery and the charger. Here $g(t)$ is the switch function, which equals to $1$ in $t\in[0,\tau]$ while zero elsewhere. The time $\tau$ represents the charging time. In our previous work,\textsuperscript{\cite{Liu21}} it has been proved that the battery performance is best in the case of $\Delta=0$. Therefore, we only consider $\Delta=0$ in this work. The charging process of the central-spin battery is shown in Fig. \ref{fig1}. 

    To maximize the energy transportation in the charging process, we set the following initial states. At time $t<0$, the battery spin is prepared in the ground state, i.e., all spins are down
        \begin{equation}
        |0\rangle_{b}\equiv|\downarrow_{1},\downarrow_{2},\cdot\cdot\cdot,\downarrow_{N_{b}}\rangle_{b}.
        \end{equation}
    While the charger spins are all prepared in the excited states
        \begin{equation}\label{Dicke state}
        |n\rangle_{c}\equiv|\uparrow_{1},\uparrow_{2},\cdot\cdot\cdot,\uparrow_{N_{c}}\rangle_{c},
        \end{equation}
    with $n=N_c$. So the total initial state is $|{\psi(0)}\rangle=|{0}\rangle_b\otimes|{n}\rangle_c$. When $N_c>N_b$, the energy of the charger spins in the initial state is larger than the energy which can be filled in the battery.\textsuperscript{\cite{Liu21}}

    Because of the flip-flop type interaction of the central-spin battery, the total spin in the $z$ direction is conserved, that is, $[J^{z}+S^{z},H]=0$. Therefore, we can reformulate the dynamics into the subspace with the same number of spins
        \begin{equation}\label{Invariant-subspace}
                \mathcal H_{n}=\{|0\rangle_{b}|n\rangle_{c},|1\rangle_{b}|n-1\rangle_{c},\cdot\cdot\cdot,|d\rangle_{b}|n-d\rangle_{c}\}.
        \end{equation}
    The parameter $d$ is defined as $d=\min\{{N_{b},N_{c}}\}$. The state $|m\rangle$ represents the equal superposition of all states with $m$-spin up, the so-called Dicke state.\textsuperscript{\cite{Dicke54}} The dimension of the subspace is $d+1$, which scales linearly with the number of spins in the battery or the charger. Note that the initial state $|\psi(0)\rangle$ is also enclosed.

    The central-spin battery Hamiltonian $H$ can be represented as a $(d+1)\times(d+1)$ matrix in the invariant subspace basis
        \begin{equation}\label{N-1}
                H=\begin{pmatrix}
                    b_0 & u_1    &                           \\
                    u_1 & b_1    & u_2     &                 \\
                        & \ddots & \ddots  & \ddots  &       \\
                        &        & u_{d-1} & b_{d-1} & u_{d} \\
                        &        &         & u_{d}   & b_{d}
                  \end{pmatrix}.
        \end{equation}
    Assuming $\omega_b = \omega_c \equiv \omega$ and $\Delta =0$, we have the matrix elements
        \begin{subequations}
            \begin{align}
                u_{j}= & j\lambda\sqrt{(N_b-j+1)(N_{c}-j+1)}, \\
                b_{j}= & \frac{\omega}{2}\sqrt{(N_b-N_c)^{2}}.
            \end{align}
        \end{subequations}
    Note that the diagonal terms $b_j$ correspond to the initial energy difference between the battery and the charger.

    To evaluate the evolution operator $U(t) = e^{-iHt}$ (with $t\in[0,\tau]$), we can diagonalize the Hamiltonian, which gives
        \begin{equation}
            U(t) = V e^{-iDt} V^\dag,
        \end{equation}
    with the eigenvalue matrix $D$. Then the charging process corresponds to the evolution
        \begin{equation}\label{N-2}
            |\psi(t)\rangle=V e^{-iDt} V^\dag |\psi(0)\rangle.
        \end{equation}
    In the invariant subspace basis, the initial state is simply $|\psi(0)\rangle = (1\ 0\ \ldots\ 0)^T$ with $T$ denoting the matrix transpose. We only concern the energy in the battery, therefore knowing the reduce density matrix of the battery suffices, which is given by
        \begin{align}\label{N-3}
                \rho_b(t)& \equiv \tr_{c}(|\psi(t)\rangle\langle\psi(t)|) \nonumber \\
                &=|\bm\psi_{1} (t)|^{2}|0\rangle_b\langle0|+\cdot\cdot\cdot+|\bm\psi_{d+1}(t)|^{2}|d\rangle_b\langle d|.
        \end{align}
    Here $\psi_{k}(t)$ is the matrix elements of $|\psi(t)\rangle$ in the invariant subspace basis.

\section{Energy transportation: analytic solution}\label{S-EAET}

 \subsection{Individual charging: $N_b=1$}

    First, consider the simplest case with $N_b = 1$, namely one battery spin charged by $N_c$ charger spins. The initial state is $|\psi (0)\rangle=|0\rangle_{b}\otimes|N_c\rangle_{c}$. The evolution state is given by
        \begin{multline}
            |\psi (t)\rangle = e^{\frac{i\left(N_c-1\right)}{2}\omega t}\Big(\cos(\sqrt{N_c}\lambda t)|0\rangle_{b} |N_c\rangle_{c} \\
            -i\sin(\sqrt{N_c}\lambda t)|1\rangle_{b} |N_c-1\rangle_{c}\Big).
        \end{multline}
    The battery has the reduced density matrix
        \begin{equation}
        \label{eq:rho_n1}
            \rho_{b}(t) = \cos^{2}(\sqrt{N_c}\lambda t)|0\rangle\langle 0 |+\sin^{2}(\sqrt{N_c}\lambda t)|1\rangle\langle 1|.
        \end{equation}
    Then the energy transported from the charger to the battery is
        \begin{align}
            \Delta E (t) \equiv \tr \left(H_{b}\rho_{b}(t)\right)-\tr\left(H_{b}\rho_{b}(0)\right).
        \end{align}
    Substitute the reduced density matrix $\rho_{b}(t)$ in Eq. (\ref{eq:rho_n1}) and the battery Hamiltonian in Eq. (\ref{eq:H_b}), then we have
        \begin{equation}
            \Delta E (t) = \omega\sin^{2}\left(\sqrt{N_{c}}\lambda t\right).
        \end{equation}
    Obviously, if the charging process stops at $\tau=\pi/( 2\lambda\sqrt{N_{c}})$, the energy transportation $\Delta E (t)$ achieves the maximal $\Delta E (\tau) = \omega$. The one battery spin is fully charged. Note that the charging time $\tau$ is proportional to $1/\sqrt{N_c}$.

    The individual charging case also suggests that we can parallel charge each battery spin with $N_c$ number of chargers. And all the battery spin can be fully charged. For $N_b$ battery spins, the total energy transport is $N_b\omega$. Although the maximal energy transportation is guaranteed in parallel charging, it does not provide any quantum advantage on charging speedup.\textsuperscript{\cite{Xiang23}} We focus on collective charging in the following sections.

\subsection{Collective charging:  $N_{b}=2$}

    \begin{figure}[t]
        \includegraphics[width=3in]{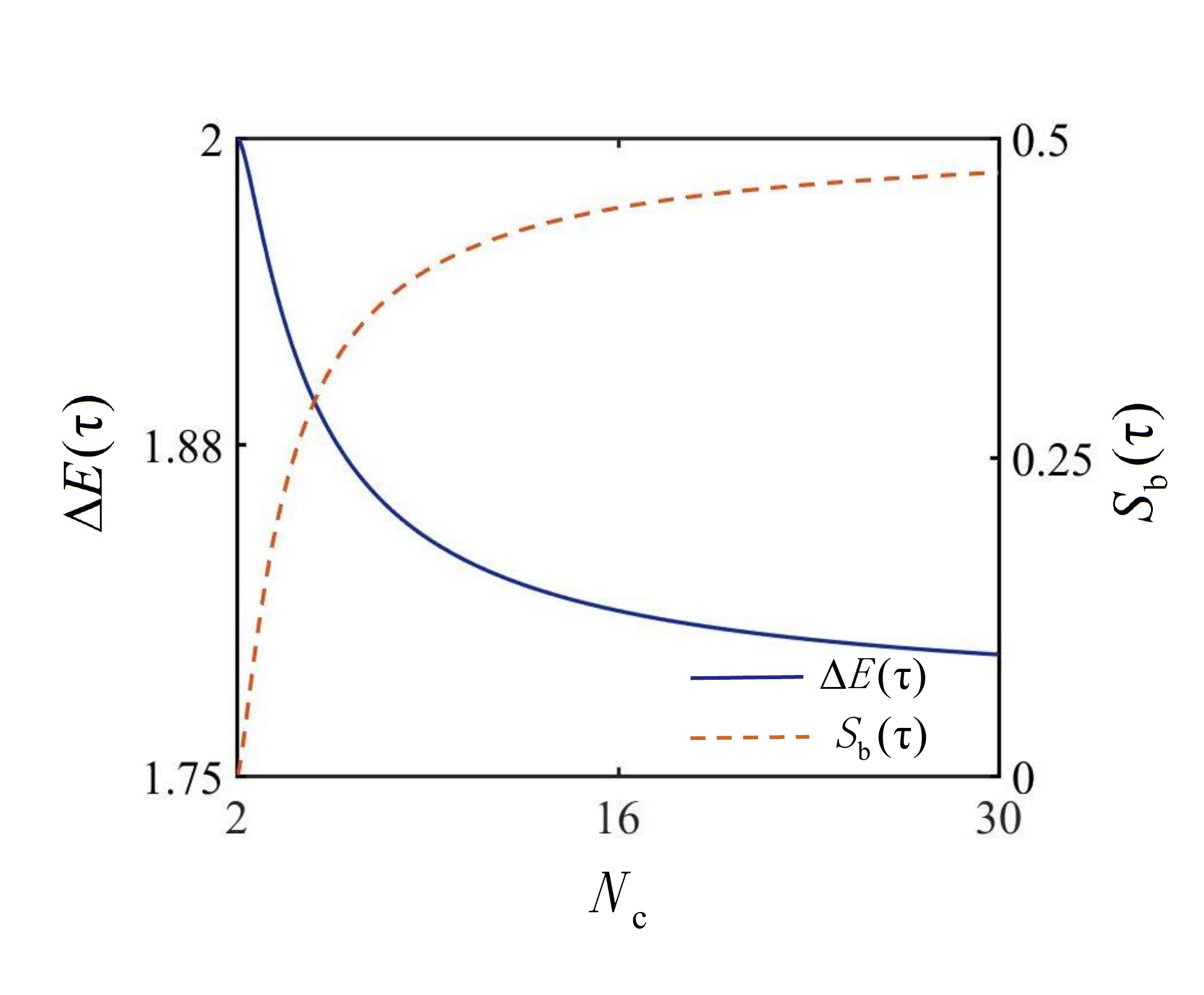}
        \caption{The battery-charger entanglement $S_b(\tau)$ and the maximal transported energy $\Delta E(\tau)$ in terms of the number of charger cells $N_c$ with $N_{b}=2\leq N_{c}$. The parameters are set to $\omega=\lambda=1$.}\label{fig2}
    \end{figure}

    Suppose that the charger spins is outnumbered the battery spins, namely $N_{b}\leq N_{c}$. If $N_b = 2$, the invariant subspace is spanned by the basis $\{|0\rangle_{b}|N_{c}\rangle_{c},|1\rangle_{b}|N_{c}-1\rangle_{c},|2\rangle_{b}|N_{c}-2\rangle_{c}\}$. Then the Hamiltonian is simply a $3\times 3$ matrix in such basis, given by
        \begin{align}
            H=\begin{pmatrix}
                      \omega(N_{c}/2-1)   & \lambda\sqrt{2N_{c}}   & 0                  \\
                      \lambda\sqrt{2N_{c}} & \omega(N_{c}/2-1)     & 2\lambda\sqrt{N_{c}-1} & \\
                      0              & 2\lambda\sqrt{N_{c}-1} & \omega(N_{c}/2-1)       \\
                  \end{pmatrix}.
        \end{align}
    Then we can analytically solve the eigenproblem. The Hamiltonian has the eigenvalues
        \begin{align}
            e_1=&\omega(N_c/2-1),\\
            e_2=&\omega(N_c/2-1)+\lambda\sqrt{2(3N_{c}-2)},\\
            e_3=&\omega(N_c/2-1)-\lambda\sqrt{2(3N_{c}-2)}.
        \end{align}
    And the corresponding transformation matrix is
        \begin{multline}
            V = \\
            \frac{1}{\sqrt{2(3N_{c}-2)}}
            \begin{pmatrix}
                2\sqrt{N_{c}-1} & \sqrt{N_{c}}      & \sqrt{N_{c}}      \\
                0               & \sqrt{3N_{c}-2}   & -\sqrt{3N_{c}-2}  \\
                -\sqrt{2N_{c}}  & \sqrt{2(N_{c}-1)} & \sqrt{2(N_{c}-1)} \\
            \end{pmatrix}.
        \end{multline}
    Then we can get the evolution state $|\psi(t)\rangle$ based on Eq. (\ref{N-2}). Tracing out the charger spins, we have the density matrix of the battery spins
        \begin{equation}
            \rho_b(t)=\rho_{11}(t)|0\rangle_b\langle0|+\rho_{22}(t)|1\rangle_b\langle1|+\rho_{33}(t)|2\rangle_b\langle2|,
        \end{equation}
    where
        \begin{subequations}
            \begin{align}\label{rho-elements}
            \begin{aligned}
                \rho_{11}(t)=&\frac{1}{\left(3N_{c}-2\right)^2}\left(2\left(N_c-1\right)+N_c \cos(\bar\omega_ct)\right)^{2},\\
                \rho_{22}(t)=&\frac{N_{c}}{3N_{c}-2}\sin^{2}\bar\omega_ct,\\
                \rho_{33}(t)=&\frac{2N_{c}(N_{c}-1)}{\left(3N_{c}-2\right)^2}\left(1-\cos(\bar\omega_ct)\right)^{2},\\
            \end{aligned}
        \end{align}
        \end{subequations}
    with $\bar\omega_c=\lambda\sqrt{2(3N_{c}-2)}$. The energy transported in the battery at time $t$ is
        \begin{multline}
        \label{energyNN2}
            \Delta E(t)=\frac{\omega}{(3N_{c}-2)^{2}}\Big(N_{c}(N_{c}-2)\cos^{2}(\bar\omega_ct)\\
            -8N_{c}(N_{c}-1)\cos(\bar\omega_ct)+N_{c}(7N_{c}-6)\Big).
        \end{multline}
    The transported energy reaches the maximal at $\tau = \pi/\bar\omega_c$. And the corresponding maximal transported energy is
        \begin{equation}\label{energyEt2}
            \Delta E(\tau)=\frac{16 \omega N_{c}(N_{c}-1)}{(3N_{c}-2)^2}.
        \end{equation}
    Note that the battery-charger interaction strength $\lambda$ is directly related to the optimal charging time $\tau$. While the maximal transported energy is proportional to $\omega$.

    When $N_c=2$, the battery with two spins can be fully charged, namely $\Delta E(\tau) = 2\omega$. However, the maximal transported energy decreases with the increasing of the charger number $N_c$. See Fig. \ref{fig2}. Asymptotically $N_c\rightarrow\infty$, we have $\Delta E(\tau) \rightarrow 16\omega/9$. Although increasing the number of charger spins can reduce the charging time in the collective charging, the battery can not be fully charged. We can also see that the diagonal terms (population) of the battery density matrix becomes
        \begin{equation}
            \rho_{11}(\tau) \rightarrow \frac{1}{9},\quad \rho_{22}(\tau) \rightarrow 0,\quad \rho_{33}(\tau) \rightarrow \frac 8 9,
        \end{equation}
    as $N_c\rightarrow\infty$.

    As the battery is not fully charged, the density matrix of battery is a mixed state. It suggests that the battery spins are entangled with the charger spins. As demonstrated in Ref. \cite{Shi22}, the energy storage of quantum battery is limited by the battery-charger entanglement. Since the quantum state of the battery-charger system remains in a pure state over the course of time evolution, the battery-charger entanglement can be well characterized by the von Neumann entropy of the battery state (or the charger state).  Specifically, the von Neumann entropy of the battery state is given by
        \begin{eqnarray}
            S_b(t)=-\tr\left(\rho_{b}(t)\log_{2}\rho_{b}(t)\right),
        \end{eqnarray}
    which also equals to the von Neumann entropy of the charger state.

    \begin{figure}[t]
        \includegraphics[width=3in]{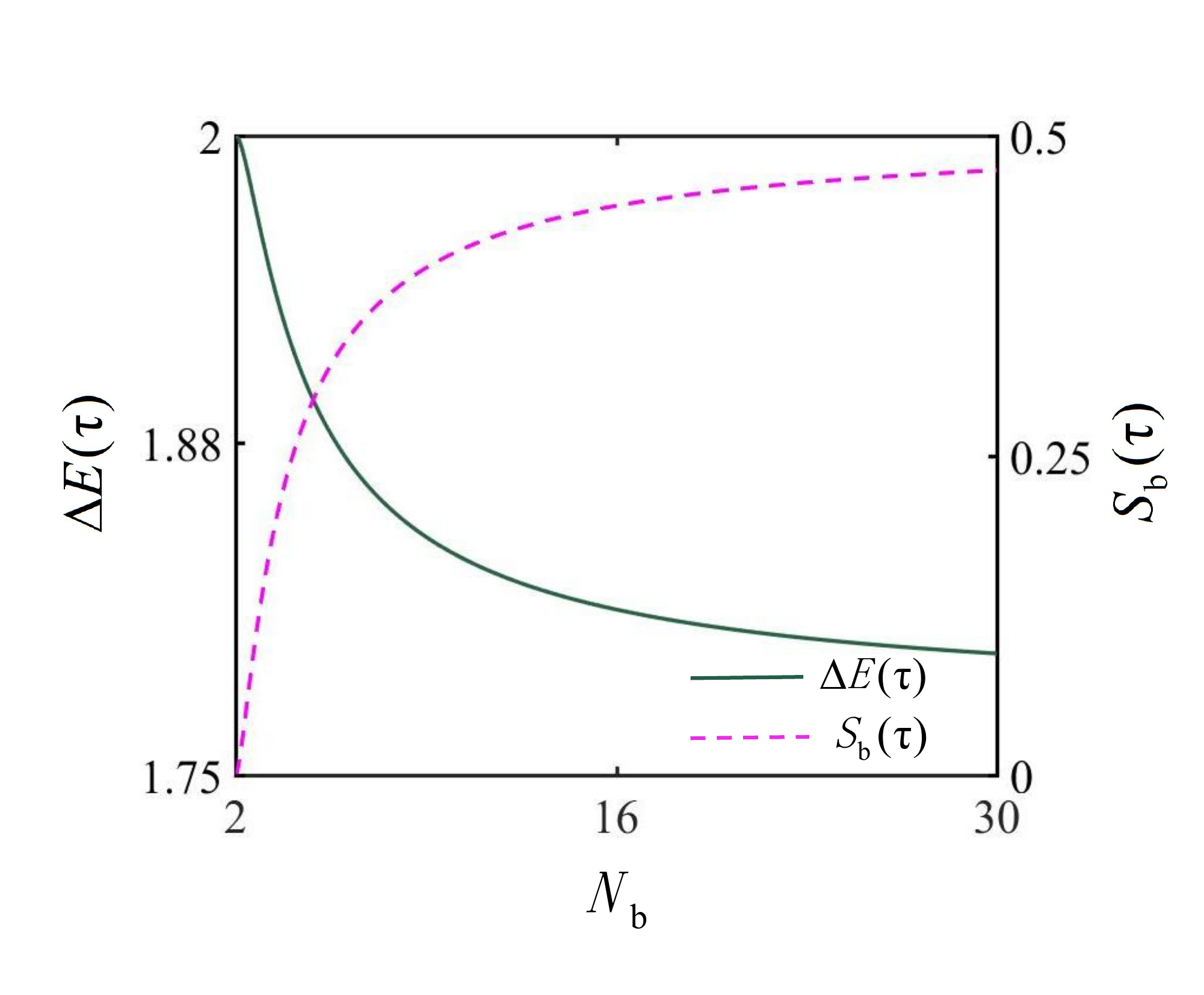}
        \caption{The battery-charger entanglement $S_b(\tau)$ and the maximal transported energy $\Delta E(\tau)$ in terms of the number of battery $N_b$ with $N_{c}=2\leq N_{b}$. The parameters are set to $\omega=\lambda=1$.}\label{fig3}
    \end{figure}

    Since the battery has a pure initial state, it has a zero von Neumann entropy $S_b(0) = 0$. For incoherent quantum batteries, entanglement is a necessary condition for generating extractable work during the charging process.\textsuperscript{\cite{Shi22}} At $\tau = \pi/\bar\omega_c$, the battery has the maximal stored energy. We find that the corresponding von Neumann entropy of the battery state is given by
        \begin{equation}\label{energySt2}
            S_b(\tau)=h\left(\left(\frac{N_{c}-2}{3N_{c}-2}\right)^2\right),
        \end{equation}
    with the binary Shannon entropy function
        \begin{equation}
        \label{eq:h(x)}
            h(x)=-x\log_{2}x-(1-x)\log_{2}(1-x).
        \end{equation}
    Since $h(x)$ is monotonic increasing with $x$ in $0\leq x \leq 1/2$, the von Neumann entropy $S_b(\tau)$ increases with $N_c$. See Fig. \ref{fig2}. Intuitively, increasing the number of charger cells would decrease the optimal charging time, which is valid in the central-spin battery. However, larger number of charger spins would generate more battery-charger entanglement, which is against the battery to be fully charged.

\subsection{Collective charging: $N_{c}=2$}

\begin{figure*}[t]
        \includegraphics[width=\textwidth]{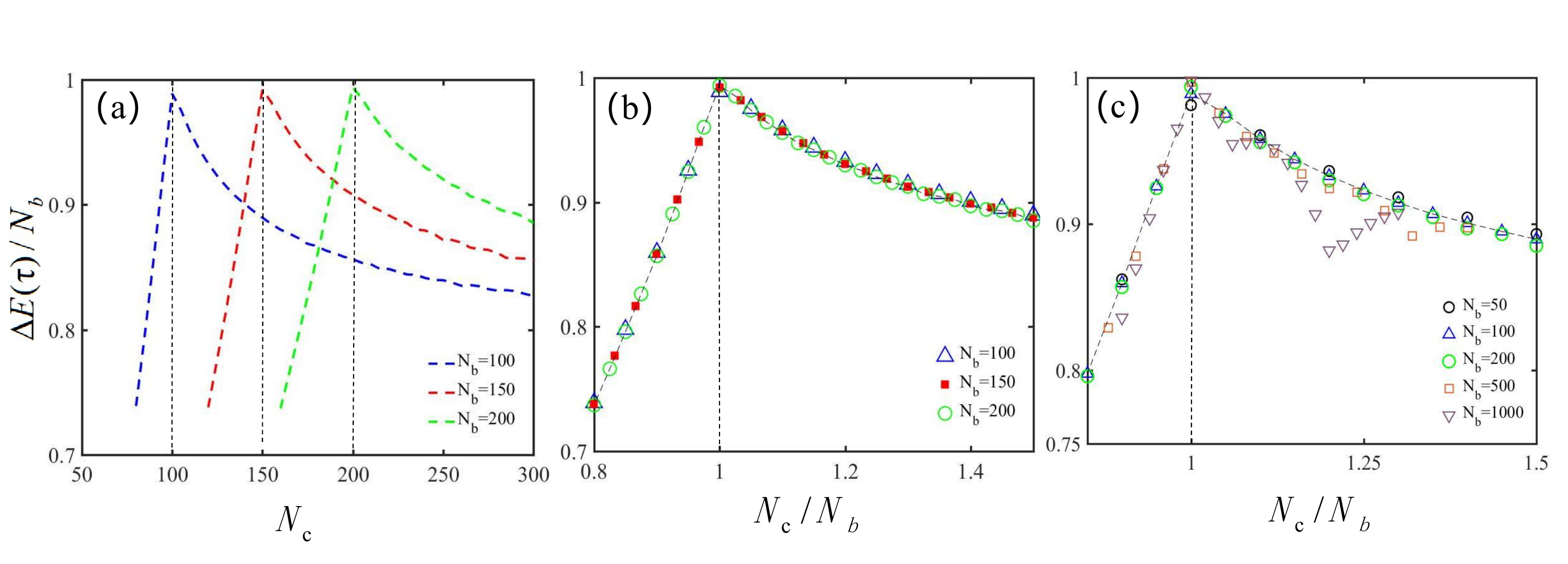}\label{S-CSB}
        \caption{Maximal transported energy per battery cell $\Delta E(\tau)$ in terms of (a) the number of charger cells or (b)(c) the battery-charger number ratio $N_c/N_b$. The parameters are set to $\omega = \lambda = 1$.}\label{fig4}
    \end{figure*}

    If there are two charger spins, namely $N_{c} = 2$, the invariant subspace is spanned by the basis $\{|0\rangle_{b}|2\rangle_{c},|1\rangle_{b}|1\rangle_{c},|2\rangle_{b}|0\rangle_{c}\}$. In other words, $N_b$ battery spins can only be double excited, while others are remained in the ground state. In this case, the Hamiltonian can also be cast into a $3\times 3$ matrix, which is analytically tractable.
    
    Following the similar procedure in the last subsection, we find that the transported energy is
        \begin{multline}
         \Delta E(t)=\frac{\omega}{(3N_{b}-2)^{2}}\Big(N_{b}(N_{b}-2)\cos^{2}(\bar\omega_bt)\\
            -8N_{b}(N_{b}-1)\cos(\bar\omega_bt)+N_{b}(7N_{b}-6)\Big)
            ,
        \end{multline}
    where $\bar\omega_b=\lambda\sqrt{2(3N_{b}-2)}$. The battery has the maximal energy at $\tau=\pi/\omega_b$. And the corresponding transported energy is
        \begin{equation}\label{energyEt3}
            \Delta E(\tau)=\frac{16 \omega N_{b}(N_{b}-1)}{(3N_{b}-2)^2}.
        \end{equation}
    The maximal transported energy $\Delta E_{\tau}$ decreases with the increasing of the number of battery cells $N_b$. Asymptotically $N_b\rightarrow\infty$, the initial charger energy $2\omega$ can not fully transport to the battery. See Fig. \ref{fig3}.

    We analyze the battery-charger entanglement quantified by the von Neumann entropy of the battery (charger) spins. At the optimal charging time $\tau$, we have the von Neumann entropy
        \begin{equation}
        \label{energySt3}
            S_b(\tau) = h \left(\left(\frac{N_{b}-2}{3N_{b}-2}\right)^2\right).
        \end{equation}
    with the binary Shannon entropy function $h(x)$ defined in Eq. (\ref{eq:h(x)}). We can see that energy transport is negatively related with the entanglement, and the battery cells can be fully charged only at $N_{b} = N_{c}$.

\section{Entanglement and energy transportation: numerical analysis}\label{S-COET}
\subsection{Uniform behaviors of energy transportation}\label{S-COT}

    In the previous section, we have established analytical results on the maximal transported energy $\Delta E(\tau)$ and the battery-charger entanglement $S_b(\tau)$ with $N_b=2$ or $N_c=2$. Although the battery-charger entanglement is necessary during the charging process in central-spin quantum battery, any entanglement after the charging process is unwanted since it is against the battery to be fully charged.\textsuperscript{\cite{Liu21}} As $N_b\gg N_c$ or $N_c\gg N_b$, the degrees of freedom of battery or charger is dominated. Therefore, it is expected that the battery and the charger become more entangled. We conjecture that the battery has the maximal transported energy per battery cell at $N_b = N_c$.

    The invariant subspace method introduced in Sec. \ref{S-CB} allows us to numerically track the dynamics with larger number of spins. First, we take the number of battery cells $N_b$ as $100,150,200$. Then calculate the maximal transported energy with different numbers of $N_c$ (by numerically finding the optimal charging time $\tau$). See Fig. \ref{fig4}(a). The maximal transported energy per battery cell, namely $\Delta E(\tau)/N_b$, is largest at $N_b = N_c$, which is consistent with our analytical results on $N_b = 2$.

    Figure \ref{fig4}(a) shows an interesting uniform behavior of the maximal transported energy. To give a fair comparison, we plot the maximal transported energy per battery cell in terms of the battery-charger ratio $N_{c}/N_{b}$. See Fig. \ref{fig4}(b). In such battery scales, namely $N_b = 100, 150, 200$,  we find that the maximal transported energy per battery cell scales almost perfectly identical.

    By further increasing the battery scale, we find that the uniform characteristics is breaking.  See Fig. \ref{fig4}(c). Oscillation appears as $N_b \geq 500$. The maximal transported energy may be directly related to the scale difference, namely $|N_b-N_c|$, rather than the scale ratio $N_c/N_b$. Nevertheless, the resonant condition $N_{b}=N_{c}$ always gives the largest transported energy in the central-spin model, which numerically verify our conjecture.

\subsection{Entanglement and energy transportation at $N_{b}=N_{c}$}\label{S-COEAET}

    \begin{figure}[t]
       \includegraphics[width=3in]{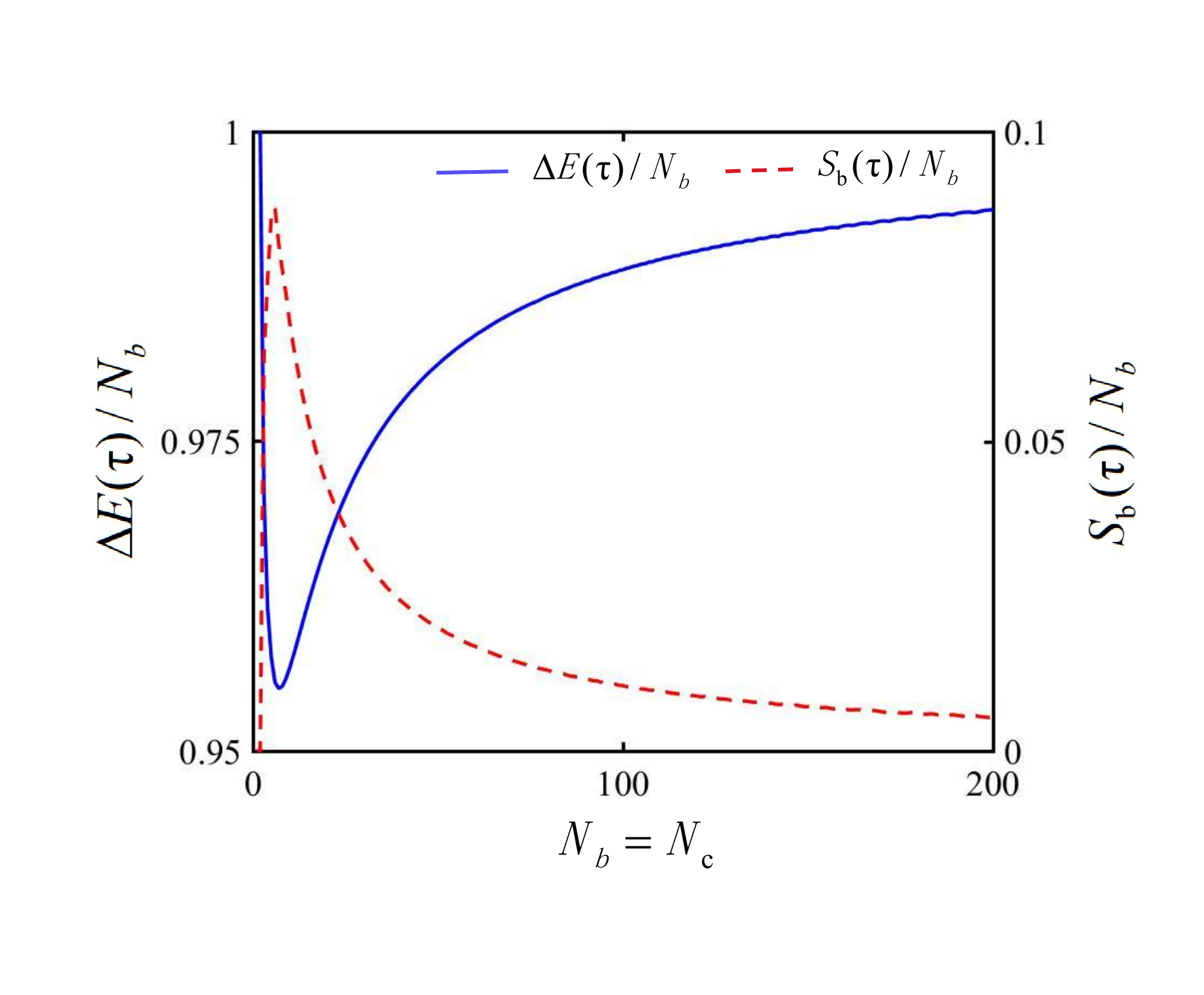}\label{S-CSB}
       \caption{Maximal transported energy per battery cell $\Delta E(\tau)$ and the battery-charger entanglement per battery cell $S_b(\tau)/N_b$ in terms of the number of battery cells $N_b$ with $N_b = N_c$. The parameters are set to $\omega = \lambda = 1$.}\label{fig5}
    \end{figure}

    In Sec. \ref{S-EAET}, we know that all the charger energy can transport to the battery when $N_b = N_c =2$, which is the most economic charging protocol. Even though we know that the battery has the maximal transported energy $\Delta E(\tau)$ at $N_b = N_c$, it does not guarantee that all the charger energy can be completely transport to the battery.

    First, we study how the size of the battery cells $N_b$ with $N_b = N_c$ influence the maximal transported energy per battery cell $\Delta E(\tau)/N_b$. See Fig. \ref{fig5}. The perfect charging only occurs at $N_b = N_c =2$. As the number of battery cells increases, the maximal transported energy per battery cell decreases. However, the energy transportation returns to the optimal as $N_b = N_c \rightarrow \infty$. We have a minimum of $\Delta E(\tau)/N_b$ around $N_b = 7$. It is expected that the battery-charger entanglement behaves inversely, namely having the minimum at $N_b = N_c =2$. Then increases with the number of battery cells. We can see that the maximal transported energy per battery cell is closely related to the battery-charger entanglement per battery cell. See Fig. \ref{fig5}.

    Based on the above results, we propose the optimal charging protocol for the central-spin battery as follows. For small size of battery, such as $N_b<200$, pair each two battery cells with two charger cells, then parallel charging the battery into full capacity. For relative large size of battery, pair all the battery cells with the same number of charger cells, then collectively charge the battery to the maximal, with the almost full capacity. Note that it is also worthwhile to apply the collective charging protocol as the battery size is large in the viewpoint of quantum charging speedup.

\section{Conclusion and outlook}\label{S-CON}

    In this work, we systematically analyse the energy transportation in central-spin battery. First, we obtain analytical results on the battery capacity and the battery-charger entanglement with $N_b =2$ or $N_c = 2$. The analytical results clearly show that the battery has the maximal capacity when $N_b = N_c = 2$. Therefore we conjecture that the resonant condition $N_b = N_c$ always gives the maximal (central-spin) battery capacity, and numerically verify it up to 2000 battery cells (spins). Second, for midsize battery, namely $N_b \sim 150$, we find a uniform relationship between the energy capacity (per charger cell) and the battery-charger size ratio $N_c/N_b$. The uniform relation for the midsize battery also gives a tight upper bound on the energy capacity when the battery cells become large. Third, we demonstrate the universal inverse relationship between the battery capacity and the battery-charger entanglement. Since the highest excited state of the battery spin, corresponding to the maximal battery storage, is a pure state, it is expected that any battery-charger entanglement prohibits the battery to reach the maximal energy capacity. It is also consistent with general theory of quantum battery.\textsuperscript{\cite{Shi22}}

    Charging process is a typical nonequilibrium dynamics, which is a challenging problem especially for the many-body system. The uniform behavior of the battery capacity may suggest that the conserved charges,\textsuperscript{\cite{Villazon20,Tang23}} which are independent on the number of spins, play a role here. The resonant condition, namely $N_b = N_c$, is a critical point for the central-spin battery. Whether a similar results exist in other types of quantum battery, such as the spin-chain quantum battery,\textsuperscript{\cite{Le18}} Dicke quantum battery,\textsuperscript{\cite{Ferrar18,Dou22,Xiang23}} and the Sachdev-Ye-Kitaev quantum battery,\textsuperscript{\cite{Rossini20}} is an open problem. We leave above questions for future study.

\section*{ACKNOWLEDGMENTS}

    This work was supported by the NSFC (Grants No. 12275215, No. 12305028, and No. 12247103), the Major Basic Research Program of Natural Science of Shaanxi Province (Grant No. 2021JCW-19), Shaanxi Fundamental Science Research Project for Mathematics and Physics (Grant No. 22JSZ005) and the Youth Innovation Team of Shaanxi Universities.


\begin{thebibliography}{99}

    \bibitem{Nielsen} Nielsen M A and Chuang I {2000 \emph{Quantum computation and quantum information }(Cambridge University Press)}

    \bibitem{Bennett93} Bennett C H, Brassard G, Popescu S, Schumacher B, Smolin G A and Wootters W K \href{https://doi.org/10.1103/PhysRevLett.76.722}{1996 \emph{Phys. Rev. Lett.} \textbf{76} 722}

    \bibitem{Gisin02} Gisin N, Ribordy G, Tittel W and Zbinden H \href{https://doi.org/10.1103/RevModPhys.74.145}{2002 \emph{Rev. Mod. Phys. }\textbf{74} 145}

    \bibitem{Ladd10} Ladd T D, Jelezko F, Laflamme R, Nakamura Y, Monroe C and O' Brien J L \href{https://doi.org/10.1038/nature08812}{2010 \emph{Nature }\textbf{465} 45-53}

    \bibitem{Reiher17} Reiher M, Wiebe N, Svore K M, Wecker D and Troyer M \href{https://www.pnas.org/doi/full/10.1073/pnas.1619152114}{ 2017 \emph{Proc. Natl. Acad. Sci. U.S.A. }\textbf{114} 7555}

    \bibitem{Rudinger22} Boixo S, Isakov S V, Smelyanskiy V N, Babbush R, Ding N,
    Jiang Z, Bremner M J, Martinis and H. Neven \href{https://doi.org/10.1038/s41567-018-0124-x}{2018 \emph{Nat. Phys.} \textbf{14} 595}

    \bibitem{Duan01} Duan L M, Lukin M D,  Cirac J I and Zoller P  \href{https://doi.org/10.1038/35106500}{2001 \emph{Nature(London) }\textbf{414} 413}

    \bibitem{Koepsell21}Koepsell J, Bourgund D, Sompet P, Hirthe S, Bohrdt A, Wang Y, Grusdt F, Demler E, Salomon G, Gross C and Bloch  \href{https://www.science.org/doi/full/10.1126/science.abe7165}{2021 \emph{Science }\textbf{374} 82}

    \bibitem{Muniz20}  Muniz J A, Barberena D, Lewis-Swan R J, Young D J, Cline J R K, Rey A M and Thompson J K \href{https://doi.org/10.1038/s41586-020-2224-x}{ 2020 \emph{Nature(London) }\textbf{580} 602}

    \bibitem{Blatt12} Blatt R and Roos C F \href{https://doi.org/10.1038/nphys2252}{ 2020 \emph{Nat. Phys. }\textbf{8} 277}

    \bibitem{Tamura20} Tamura M,  Mukaiyama T and Toyoda K \href{https://doi.org/10.1103/PhysRevLett.124.200501}{ 2020 \emph{Phys. Rev. Lett. }\textbf{124} 200501}

    \bibitem{Niu21} Niu J J, Yan T X, Zhou Y X, Tao Z Y, Li X L,  Liu W Y, Zhang L B, Jia H, Liu S, Yan Z B, Chen Y Z, Yu D \href{https://doi.org/10.1016/j.scib.2021.02.035}{2021 \emph{Sci Bull} \textbf{66} 1168}

    \bibitem{Guo21} Guo Q J, Cheng C, Sun Z H, Song Z X,  Li H K, Wang Z, Ren W H, Dong H, Zheng D N, Zhang Y R, Mondaini R, Fan H and Wang H \href{https://doi.org/10.1038/s41567-020-1035-1}{2021 \emph{Nat. Phys.} \textbf{17} 234}

    \bibitem{Campisi11} Campisi M, Hanggi P and Talkner P \href{https://doi.org/10.1103/RevModPhys.83.1653}{ 2011 \emph{Rev. Mod. Phys.} \textbf{83} 1653}

    \bibitem{Yang23} Yang X, Yang Y H, Alimuddin M,  Salvia R, Fei S M, Zhao L M, Nimmrichter S and  Luo M X \href{https://doi.org/10.1103/PhysRevLett.131.030402}{ 2023 \emph{Phys. Rev. Lett.} \textbf{131} 030402}

    \bibitem{shi20} Shi Y H, Shi H L, Wang X H, Hu M L, Liu S Y, Yang W L and Fan H \href{https://doi.org/10.1088/1751-8121/ab6a6b}{ 2020 \emph{Journal of Physics A: Mathematical and Theoretical} \textbf{53} 085301}

    \bibitem{Ji22} Ji W, Chai Z, Wang M, Guo Y, Rong X, Shi F, Ren C L, Wang Y and Du J F \href{https://doi.org/10.1103/PhysRevLett.128.090602}{ 2022 \emph{Phys. Rev. Lett.} \textbf{128} 090602}

    \bibitem{Dou22} Dou F Q, Lu Y Q, Wang Y J and Sun J A  \href{https://doi.org/10.1103/PhysRevB.105.115405}{ 2022 \emph{Phys. Rev. B} \textbf{105} 115405}

    \bibitem{Dou22A} Dou F Q, Zhou H and Sun J A \href{https://doi.org/10.1103/PhysRevA.106.032212}{ 2022 \emph{Phys. Rev. A} \textbf{106} 032212}

    \bibitem{Wang20} Wang Z, Li H, Feng W, Song X, Song C, Liu W, Guo Q, Zhang X, Dong H, Zheng D, Wang H and Wang D W \href{https://dx.doi.org/10.1103/PhysRevLett.124.013601}{ 2022 \emph{Phys. Rev. Lett.} \textbf{124} 013601}

    \bibitem{Lu21} Lu W, Chen J, Kuang L M and Wang X \href{https://doi.org/10.1103/PhysRevA.104.043706}{ 2021 \emph{Phys. Rev. A} \textbf{104} 043706}

    \bibitem{Uzdin15} Uzdin R, Levy A and Kosloff R \href{https://doi.org/10.1103/PhysRevX.5.031044}{ 2015 \emph{Phys. Rev. X} \textbf{5} 031044}

    \bibitem{F13} Brand$\tilde{a}$o F G S L, Horodecki M, Oppenheim J, Renes J M and Spekkens R W  \href{https://doi.org/10.1103/PhysRevLett.111.250404}{ 2013 \emph{Phys. Rev. Lett.} \textbf{111} 250404}

    \bibitem{Barrios17} Alvarado Barrios G, Albarr$\acute{a}$n-Arriagada F, C$\acute{a}$rdenas-L$\acute{o}$pez F A, Romero G and Retamal J C \href{https://doi.org/10.1103/PhysRevA.96.052119}{ 2017 \emph{Phys. Rev. A} \textbf{996} 052119}

    \bibitem{Altintas14} Altintas F, Hardal A $\ddot{U}$ C and M$\ddot{u}$stecaplio$\breve{g}$lu $\ddot{O}$ E \href{ https://doi.org/10.1103/PhysRevE.90.032102}{ 2014 \emph{Phys. Rev. E} \textbf{90} 032102}

    \bibitem{Park13} Park J J, Kim K H, Sagawa T and Kim S W \href{https://doi.org/10.1103/PhysRevLett.111.230402}{ 2017 \emph{Phys. Rev. Lett.} \textbf{111} 230402}

    \bibitem{Seah21} Seah S, Perarnau-Llobet M, Haack G, Brunner N and Nimmrichter S \href{https://doi.org/10.1103/PhysRevLett.127.100601}{ 2017 \emph{Phys. Rev. Lett.} \textbf{127} 100601}

    \bibitem{Manzano18} Manzano G, Plastina F and Zambrini R \href{https://doi.org/10.1103/PhysRevLett.121.120602}{ 2017 \emph{Phys. Rev. Lett.} \textbf{121} 120602}

    \bibitem{Goold16} Goold J, Huber M, Riera A, del Rio L and Skrzypczyk P \href{https://doi.org/10.1088/1751-8113/49/14/143001}{2016
    \emph{J. Phys. A} \textbf{49} 143001}

    \bibitem{Strasberg17} Strasberg P, Schaller G, Brandes T and Esposito M  \href{https://doi.org/10.1103/PhysRevX.7.021003}{ 2017
    \emph{Phys. Rev. X} \textbf{7} 021003}

    \bibitem{Watanabe17} Watanabe G, Venkatesh B P, Talkner P and del Campo A \href{https://doi.org/10.1103/PhysRevLett.118.050601}{2017
    \emph{Phys. Rev. Lett.} \textbf{118}, 050601}

    \bibitem{Zhang19} Zhang Y Y, Yang T R, Fu L and Wang X \href{https://doi.org/10.1103/PhysRevA.107.022209}{2019 \emph{Phys. Rev. E} \textbf{99} 052106}

    \bibitem{Alicki13} Alicki R and Fannes M \href{https://doi.org/10.1103/PhysRevE.87.042123}{2013
    \emph{Phys. Rev. E} \textbf{87} 042123}

    \bibitem{Quach22} Quach J Q, McGhee K E, Ganzer L, Rouse D M, Lovett B W, Gauger E M, Keeling J, Cerullo G, Lidzey D G and Virgili T
    \href{https://doi.org/10.1126/sciadv.abk3160}{2022 \emph{Sci. Adv.} \textbf{8} eabk3160}
    
    \bibitem{Liu21} Liu J X, Shi H L, Shi Y H, Wang X H and Yang W L \href{https://doi.org/10.1103/PhysRevB.104.245418}{2021 \emph{Phys. Rev. B} \textbf{104} 245418}

    \bibitem{Ferrar18} Ferraro D, Campisi M, Andolina G M, Pellegrini V and Polini M \href{https://doi.org/10.1103/PhysRevLett.120.117702}{ 2018 \emph{Phys. Rev. Lett.} \textbf{120} 117702}

    \bibitem{Fusco16} Fusco L, Paternostro M and Chiara G D \href{https://doi.org/10.1103/PhysRevE.94.052122}{ 2016 \emph{Phys. Rev. E} \textbf{94} 052122}

    \bibitem{Binder15} Binder F C, Vinjanampathy S, Modi K and Goold J  \href{https://doi.org/10.1088/1367-2630/17/7/075015}{2015
    \emph{New J. Phys.} \textbf{17} 075015}

    \bibitem{Andolina18} Andolina G M, Farina D, Mari A, Pellegrini V, Giovannetti V and Polini M  \href{https://doi.org/10.1103/PhysRevB.98.205423}{ 2018 \emph{Phys. Rev. B} \textbf{98} 205423}

    \bibitem{Santos21} Santos A C \href{https://doi.org/10.1103/PhysRevE.103.042118}{2021 \emph{Phys. Rev. E} \textbf{103} 042118}

    \bibitem{Rossini20} Rossini D, Andolina G M, Rosa D, Carrega M and Polini M \href{https://doi.org/10.1103/PhysRevLett.125.236402}{2020 \emph{Phys. Rev. Lett.} \textbf{125} 236402}

    \bibitem{Yu23} Yu W L, Zhang Y, Li H, Wei G F, Han L P, Tian F and Zou J
    \href{https://doi.org/10.1088/1674-1056/ac728b}{2023 \emph{Chin. Phys. B} \textbf{32} 010302}
    
    \bibitem{Yang231} Yang Z Q, Zhou L K, Zhou Z Y, Jin G R, Cheng L and Wang X G
    \href{https://doi.org/10.1088/1674-1056/acdc12}{2023 \emph{Chin. Phys. B} \textbf{32} 110301}

    \bibitem{Yao21} Yao Y and Shao X Q \href{https://doi.org/10.1103/PhysRevE.104.044116}{2021 \emph{Phys. Rev. E} \textbf{104} 044116}

    \bibitem{Andolina19} Andolina G M, Keck M, Mari A, Campisi M, Giovannetti V and Polini M \href{https://doi.org/10.1103/PhysRevLett.122.047702}{ 2019 \emph{Phys. Rev. Lett.} \textbf{122} 047702}

    \bibitem{Xiang23} Zhang X and Blaauboer M \href{https://doi.org/10.3389/fphy.2022.1097564}{2023 \emph{Front. Phys.} \textbf{10} 1097564}

    \bibitem{Jian07} Li J and Shen S Q \href{https://doi.org/10.1103/PhysRevB.76.153302}{2007 \emph{Phys. Rev. B} \textbf{76} 153302}

    \bibitem{Bortz20} Bortz M and Stolze J \href{https://doi.org/10.1103/PhysRevB.76.014304}{2007
    \emph{Phys. Rev. B} \textbf{76} 014304}

    \bibitem{Gaudin76} Gaudin M \href{https://doi.org/10.1051/jphys:0197600370100108700}{1976 \emph{J. Phys.} \textbf{37} 10}

    \bibitem{Gaudin22} Dominguez F, Esebbag C and Dukelsky J \href{https://doi.org/10.1088/0305-4470/39/37/002}{2006 \emph{J. Phys. A} \textbf{39} 11349}

    \bibitem{Faribault19} Faribault A, Koussir H and Mohamed M H \href{https://doi.org/10.1103/PhysRevB.100.205420}{2019 \emph{Phys. Rev. B} \textbf{100} 205420}

    \bibitem{Doherty13} Doherty M W, Manson N B, Delaney P, Jelezko F, Wrachtrup J and Hollenberg L C L \href{https://doi.org/10.1016/j.physrep.2013.02.001}{2013 \emph{Phys. Rep.} \textbf{528} 1}

    \bibitem{JY23} Fan J Y and Pang S S \href{https://doi.org/10.1103/PhysRevA.107.022209}{2023 \emph{Phys. Rev. A} \textbf{107} 022209}

    \bibitem{Li20} Li Z, Yang P, You W L and Wu N \href{https://doi.org/10.1103/PhysRevA.102.032409}{2023 \emph{Phys. Rev. A} \textbf{102} 032409}

    \bibitem{Schliemann02} Schliemann J, Khaetskii A V and Loss D \href{https://doi.org/10.1103/PhysRevB.66.245304}{2002 \emph{Phys. Rev. B} \textbf{66} 245304 }

    \bibitem{Khaetskii02} Alexander V. Khaetskii, Loss D and Glazman L \href{https://doi.org/10.1103/PhysRevLett.88.186802}{2002 \emph{Phys. Rev. Lett.} \textbf{88} 186802}.

    \bibitem{Deng06} Deng C and Hu X  \href{https://doi.org/10.1103/PhysRevB.73.241303}{2006 \emph{Phys. Rev. B} \textbf{73} 241303}

    \bibitem{Peng21} Peng L, He W B, Chesi S, Lin H Q and Guan X W \href{https://doi.org/10.1103/PhysRevA.103.052220}{ 2021 \emph{Phys. Rev. A} \textbf{103} 052220}

    \bibitem{Kamin21} Kamin F H, Tabesh F T, Salimi S and Santos A C \href{https://doi.org/10.1103/PhysRevE.102.052109}{2020 \emph{Phys. Rev. E} \textbf{102} 052109}

    \bibitem{Hovhannisyan13} Hovhannisyan K V, Perarnau-Llobet M, Huber M and Acin A \href{https://doi.org/10.1103/PhysRevLett.111.240401}{2013 \emph{Phys. Rev. Lett.} \textbf{111} 240401}

    \bibitem{Shi22} Shi H L, Ding S, Wan Q K, Wang X H and Yang W L \href{https://doi.org/10.1103/PhysRevLett.129.130602}{ 2022 \emph{Phys. Rev. Lett.} \textbf{129} 130602}

    \bibitem{Dicke54} Dicke R H, \href{https://doi.org/10.1103/PhysRev.93.99}{ 1954 \emph{Phys. Rev.} \textbf{93} 99}

    \bibitem{Villazon20} Villazon T, Chandran A and Claeys P W \href{https://doi.org/10.1103/PhysRevResearch.2.032052}{2020 \emph{Phys. Rev. R} \textbf{2} 032052}

    \bibitem{Tang23} Tang L H, Long D M, Polkovnikov A, Chandran A and Claeys P W \href{https://www.scipost.org/SciPostPhys.15.1.030}{2023 \emph{Scipost Phys.} \textbf{15} 030}

    \bibitem{Le18} Le T, Levinsen J, Modi K, Parish M M and Pollock F A
    \href{https://www.scipost.org/SciPostPhys.15.1.030}{2018 \emph{Phys. Rev. A} \textbf{97} 022106}



\end{thebibliography}
\end{document}